\documentclass[prd,superscriptaddress,floatfix,amssymb]{revtex4}
\usepackage{epsf}
\newcommand{\picdir}[1]{./#1}     

\def\lesssim{\mathrel{\hbox{\rlap{\hbox{\lower4pt\hbox{$\sim$}}}\hbox{$<$}}}}
\def\gtrsim{\mathrel{\hbox{\rlap{\hbox{\lower4pt\hbox{$\sim$}}}\hbox{$>$}}}}
\begin{document}
\title{Inflaton Fragmentation After $\lambda \phi^4$ Inflation}
\date{\today}
\author{Gary N. Felder}
\affiliation{Department of Physics, Clark Science Center, 
Smith College Northampton, MA 01063, USA}
\author{Olga Navros}
\affiliation{Department of Physics and Astronomy, University of North
Carolina-Chapel Hill, Chapel Hill, NC  27599}
\preprint{hep-ph/xxxxxxx}
\begin{abstract}
We use lattice simulations to examine the detailed dynamics of
inflaton fragmentation during and after preheating in $\lambda \phi^4$
chaotic inflation. The dynamics are qualitatively similar to
preheating after $m^2 \phi^2$ inflation, involving the exponential
growth and subsequent expansion and collision of bubble-like
inhomogeneities of the inflaton and other scalar fields. During this
stage fluctuations of the fields become strongly non-Gaussian. In the
quartic theory, the conformal nature of the theory allows us to extend
our simulations to much greater times than is possible for the
quadratic model. With these longer simulations we have been able to
determine the time scale on which Gaussianity is restored, which
occurs after a time on the order of a thousand inflaton oscillations.
\end{abstract}
\maketitle

\section{Introduction}

After inflation the energy density of the nearly homogeneous inflaton
field decays into fluctuations of the inflaton and other fields. In
many simple models of inflation this process begins with an
exponentially rapid stage of decay that produces highly inhomogeneous,
nonthermal field fluctuations. In large field chaotic inflation models
preheating occurs via parametric resonance, which has been analyzed
both analytically (see e.g. \cite{KLS}) and numerically
\cite{KT}. This preheating stage is followed by a short, violent
transition that leads to a regime of Kolmogorov turbulence \cite{MT}.

In \cite{bubbles} this transition stage was studied for a model with
inflaton potential $V = (1/2) m^2 \phi^2$. It was found there that
parametric resonance leads to growth of fluctuations in the peaks of
the initial random gaussian field, giving rise to a quasi-stable
standing wave pattern of bubbles and nodes. This growth continues
until backreaction makes parametric resonance inefficient, after which
these bubbles expand and collide, thus bringing the entire space into
the strongly inhomogeneous regime. It was found there that the
fluctuations produced during preheating are strongly non-Gaussian, and
that this non-Gaussianity persists long after the end of parametric
resonance.

For technical reasons discussed below, the simulations performed in
\cite{bubbles} could not be continued long enough to determine the
ultimate fate of these non-Gaussian perturbations. In this paper we
explore these same questions in a quartic model with an inflaton
potential $V = (1/4) \lambda \phi^4$. In this model it is possible to
run the simulations to much later times and we were thus able to see
the field statistics return to Gaussianity.

In section \ref{model} we describe the model we are using and our
numerical simulations. In section \ref{results} we present the results
of our simulations and discuss their implications.

\section{The Model}\label{model}

We consider the potential
\begin{equation}
\label{potential}
V = {1 \over 4} \lambda \phi^4 + {1 \over 2} g^2 \phi^2 \chi^2
\end{equation}
where $\phi$ is the inflaton and $\chi$ is another scalar field that
is coupled to it. We use LATTICEEASY \cite{FT} to solve the
classical equations of motion for the two fields
\begin{eqnarray}
\ddot{\phi} + 3 H \dot{\phi} - {1 \over a^2} \nabla^2 \phi +
\left(\lambda \phi^2 + g^2 \chi^2\right) \phi &=& 0 \\
\ddot{\chi} + 3 H \dot{\chi} - {1 \over a^2} \nabla^2 \chi +
g^2 \phi^2 \chi &=& 0.
\end{eqnarray}
Oscillations of the zeromode of $\phi$ after inflation lead to
parametrically resonant amplification of modes of $\chi$ within
certain resonance bands. These amplified modes of $\chi$ in turn
excite fluctuations of $\phi$.

LATTICEEASY uses a comoving grid, meaning the wavelength of produced
fluctuations remains constant in program coordinates. In a quadratic
model this comoving growth causes a problem for compuater simulations
because the physical wavelength $m^{-1}$ at which modes are
preferentially produced shrinks in these comoving variables. As the
universe expands the grid spacing needs to remain small enough to
accurately capture this physical length scale. Once the universe has
expanded enough to violate this criterion, it is no longer possible to
continue the simulation with any accuracy.

In the quartic model, however, the parameters $\lambda$ and $g$ are
unitless and there is no fixed physical length scale in the
problem. By defining a new set of variables $\phi_{pr} \equiv \phi/a$
and $\tau \equiv a t$ the equations of motion can be recast as
\begin{eqnarray}
{\phi_{pr}''} - \nabla^2 \phi_{pr} +
\left(\lambda \phi_{pr}^2 + g^2 \chi_{pr}^2\right) \phi_{pr} + \Delta &=& 0 \\
{\chi_{pr}''} - \nabla^2 \chi_{pr} +
g^2 \phi_{pr}^2 \chi_{pr} + \Delta &=& 0
\end{eqnarray}
where primes represent differentiation with respect to $\tau$ and
$\Delta$ represents derivatives of the scale factor that vanish for a
radiation equation of state, an approximation that is very accurate
for this model. \footnote{LATTICEEASY does not make this approximation
and calculates all terms, but tests with and without these terms
confirm that they are irrelevant to the evolution of the fields.} In
other words we can scale expansion out of the equations, and can thus
simulate to much later times than would be possible for a quadratic
model.

The simulations shown in this paper were done on a three dimensional
grid of $128^3$ or $256^3$ gridpoints. The simulations start at the
end of inflation when the mean value of the inflaton is $\phi_0 =
0.342 M_p$. Times are reported in units of $1/(\sqrt{\lambda} \phi_0)$
and field values are reported in units of $\phi_0$. All of the results
shown in this paper are for $\lambda = 9 \times 10^{-14}$ and
$g^2/\lambda = 200$.

\section{Inflaton Fragmentation: Results and Conclusions}\label{results}

\begin{figure}[htb]
\begin{minipage}[t]{7.5cm}
\centering \leavevmode \epsfxsize=7.5cm
\epsfbox{\picdir{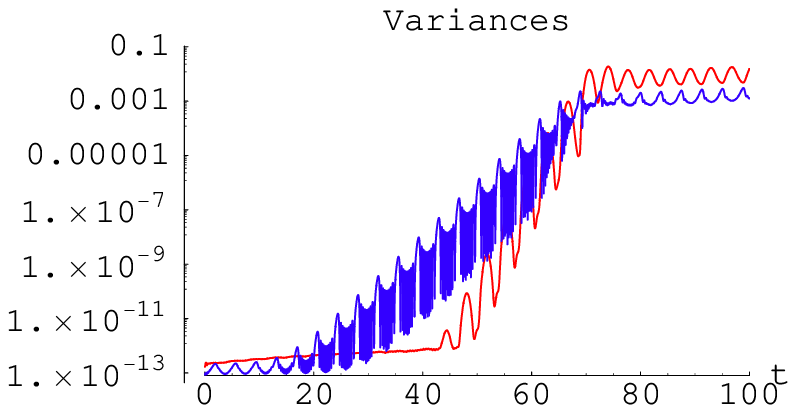}}\\
\caption{Variances of the fields $\phi$ (red) and $\chi$ (blue). (For
black and white viewing, $\chi$ begins growing earlier; $\phi$ begins
later but quickly catches up to $\chi$.)}
\label{variance}
\end{minipage}
\hspace{0.2cm}
\begin{minipage}[t]{7.5cm}
\centering \leavevmode \epsfxsize=7.5cm
\epsfbox{\picdir{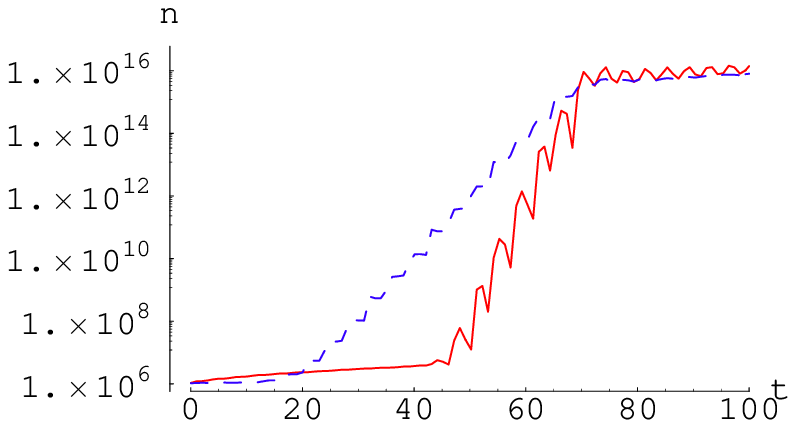}}\\
\caption{Number density for $\phi$ (red, solid) and $\chi$ (blue,
dashed).}
\label{number}
\end{minipage}
\end{figure}

Figures \ref{variance} and \ref{number} show the growth of
fluctuations of the fields $\phi$ and $\chi$. The information in these
plots is well known. The field $\chi$ grows exponentially. Some time
after the growth of $\chi$ starts fluctuations of the field $\phi$
begins growing with twice the exponent of $\chi$. These plots are
included here primarily as a reference for seeing where in the process
of preheating the fields are at each of the times shown in the plots
below.

\begin{figure}[htb]
\leavevmode\epsfxsize=.48\columnwidth\epsfbox{\picdir{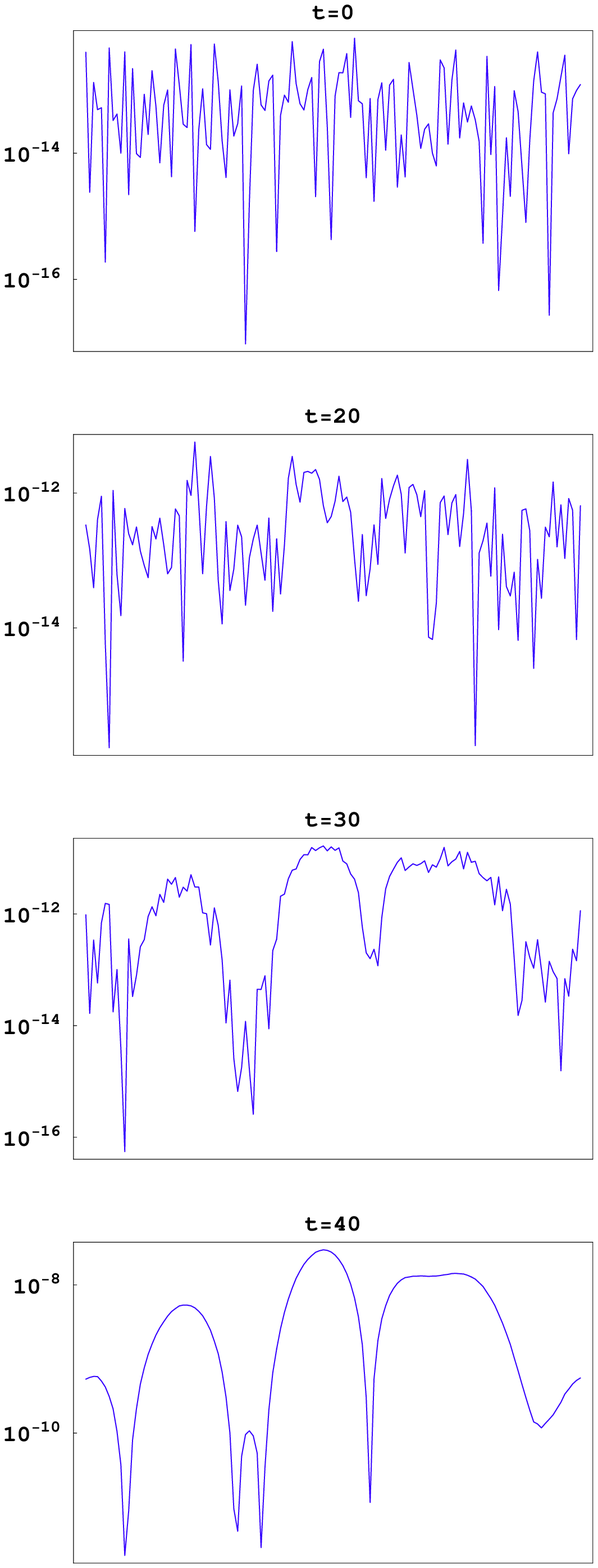}}
\hspace{.1cm}
\vline
\hspace{.1cm}
\leavevmode\epsfxsize=.48\columnwidth \epsfbox{\picdir{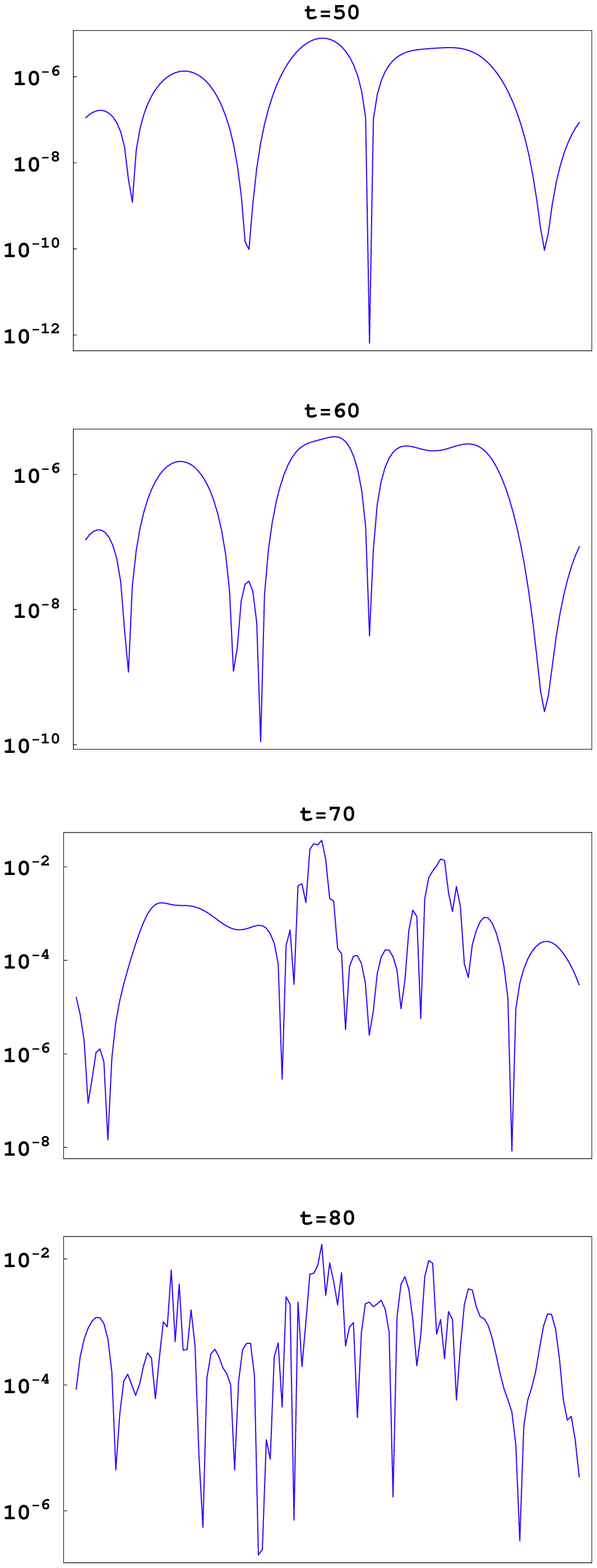}}
\caption{Values of the field $\chi$ on a 1D slice through the lattice.}
\label{peaks}
\end{figure}

Figure \ref{peaks} shows the growth of fluctuations of the $\chi$
field during preheating. For clarity the figures show values only on a
one dimensional slice through the lattice. The initial conditions used
for the model are gaussian random fluctuations with expectation values
\begin{equation}
\langle \vert \phi_k \vert^2 \rangle = {1 \over 2 \sqrt{k^2 + m_{eff}^2}},
\end{equation}
intended to simulate quantum vacuum fluctuations
\cite{KT,PS}. See the LATTICEEASY documentation
\cite{latticeeasyweb} for more details. As parametric resonance begins
fluctuations of the field begin to grow exponentially. Note that the
vertical scale on the different frames of \ref{peaks} is not
constant. Since parametric resonance only excites modes with momenta
below a certain cutoff ($k^* \sim (g^2 \lambda)^{1/4} \phi_0$
\cite{GKLS}), the short wavelength fluctuations rapidly become
insignificant. The longer wavelength modes, however, remain almost
entirely unchanged except for their overall scale. In other words the
spatial distribution of the fluctuations produced during preheating
simply mimics the spatial distribution of the infrared modes of $\chi$
that were present before preheating.

\begin{figure}[htb]
\leavevmode\epsfxsize=.48\columnwidth\epsfbox{\picdir{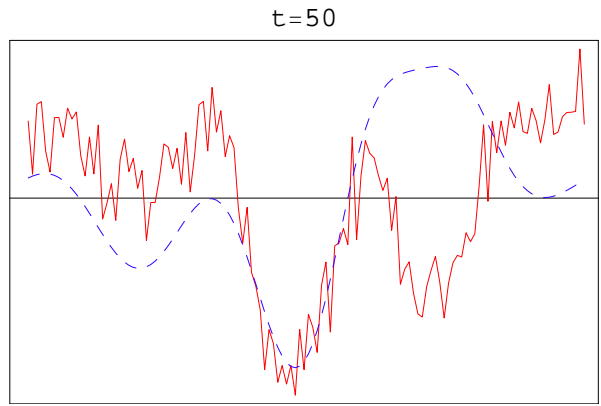}}
\caption{Values of field fluctuations $\delta \phi$ (red, solid) and
$\delta \chi$ (blue, dashed) on a 1D slice through the
lattice. Because the scale of $\delta \chi$ is much larger than
$\delta \phi$ at the time shown, fluctuations of both fields are
normalized to the same standard deviation.}
\label{phichipeaks}
\end{figure}

Fluctuations of $\phi$ are not produced until later, but when they
appear they grow due to interactions with the amplified $\chi$
fluctuations, so the peaks of $\delta \phi$ mostly correspond to the
peaks of the initial random gaussian field $\chi$. Figure
\ref{phichipeaks} shows a snapshot of $\phi$ and $\chi$ fluctuations
shortly after the $\phi$ fluctuations have started to grow. The
fluctuations of $\phi$ are for the most part in the same places as the
fluctuations of $\chi$; the correlation between $\delta \phi^2$ and
$\delta \chi^2$ in this plot is 0.72. However, the oscillation
frequencies are different for the two fields so the fluctuations are
in general not in phase, as can be seen in the plot.

\begin{figure}[htb]
\leavevmode\epsfxsize=.9\columnwidth\epsfbox{\picdir{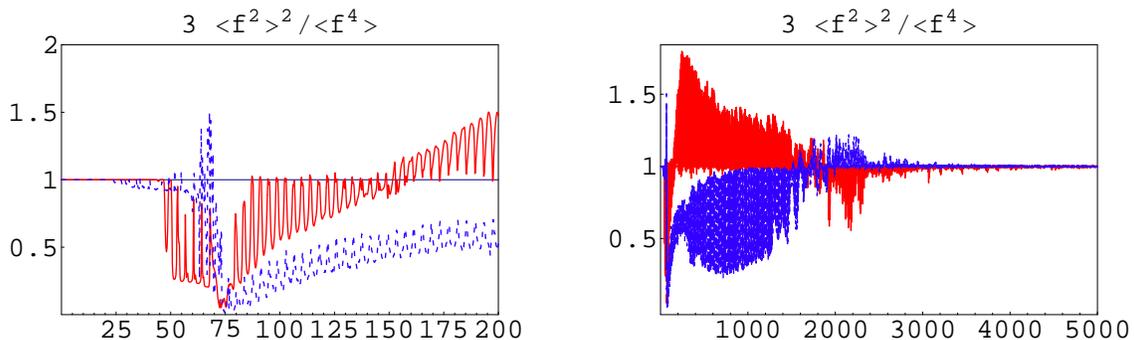}}
\caption{Kurtosis for the fields $\phi$ (red, solid) and $\chi$ (blue,
dashed). The plot on the left shows the development of non-Gaussianity
during preheating. The plot on the right shows a much larger time
scale on which the return to Gaussianity can be seen.}
\label{gaussianity}
\end{figure}

Finally, we considered the field statistics for these field
fluctuations. It is known that fluctuations produced during preheating
are non-Gaussian (see e.g. \cite{KLS}). We measure the Gaussianity of
the fields through the kurtosis, $3 \langle f^2\rangle^2/\langle
f^4\rangle$. It is a necessary but not sufficient condition for
Gaussianity that this quantity be equal to one. Figure
\ref{gaussianity} shows the evolution of this quantity for fields
$\phi$ and $\chi$. As noted in \cite{bubbles} the fluctuations become
non-Gaussian during preheating and begin slowly tending towards
Gaussianity thereafter. We see here that Gaussianity is restored after
a time of order of a thousand inflaton oscillations.

We can understand this timing with a rough analytical
estimate. Perturbatively the time scale for rescattering is given by
$t_r \sim 1/(n \sigma)$ where $n$ and $\sigma$ are the number density
and scattering cross sections for the gas of particles. The cross
section can be estimated as $\sigma \sim g^2/k^{*2}$, where $k^* \sim
(g^2 \lambda)^{1/4} \phi_0$ is the typical particle momentum produced
during preheating. To estimate the number density we can take $n \sim
\rho/k^* \sim g^2 <\phi^2> <\chi^2>/k^*$. Putting all of this together
\begin{equation}
\sqrt{\lambda} \phi_0 t_r \sim {1 \over (g^2/\lambda)^{5/4}
\langle\left(\phi/\phi_0\right)^2\rangle
\langle\left(\chi/\phi_0\right)^2\rangle}.
\end{equation}
Shortly after preheating the variances
$\langle\left(\phi/\phi_0\right)^2\rangle$ and
$\langle\left(\chi/\phi_0\right)^2\rangle$ are both approximately
$10^{-3}$, so the rescattering time is of order
\begin{equation}
\sqrt{\lambda} \phi_0 t_r \sim {1 \over (200)^{5/4} 10^{-6}} \sim 1000.
\end{equation}
In other words the time that we found is required to restore
Gaussianity has the expected order of magnitude.

In a lattice simulation all classical dynamics are automatically taken
into account, so the Gaussianity of the fields is irrelevant to the
accuracy of the simulation. While lattice simulations are excellent
for describing preheating and the evolution shortly afterwards, they
can not be extended to time scales long enough to describe the stages
of turbulence and thermalization. Many approximation techniques are
thus employed to describe these epochs (see e.g. \cite{MT} and
references therein), and for these it is often important to know the
field statistics. Our results suggest that techniques that assume
Gaussianity should not be employed immediately after preheating, but
they also suggest that it can be possible to extend lattice
simulations long enough to get through the non-Gaussian stage and thus
overlap with the subsequent period that can be well described in other
ways.

\bigskip

We would like to thank Lev Kofman for useful discussions and
advice. This work was supported by NSF grant PHY-0456631.

\end{document}